# Spin-orbit effective fields in Pt/GdFeCo bilayers

Woo Seung Ham[1], Sanghoon Kim[1], Duck-Ho Kim[1], Kab-Jin Kim[1,2], Takaya Okuno[1], Hiroki Yoshikawa[3], Arata Tsukamoto[3], Takahiro Moriyama[1]*, and Teruo Ono[1]†

[1.] Institute for Chemical Research, Kyoto University, Uji, Kyoto 611-0011, Japan

[2.] Department of Physics, Korea Advanced Institute of Science and Technology, Daejeon 34141, Korea

[3.] College of Science and Technology, Nihon University, Funabashi, Chiba 274-8501, Japan

Correspondence to: *mtaka@scl.kyoto-u.ac.jp, †ono@scl.kyoto-u.ac.jp

**Abstract**

In the increasing interests on spin-orbit torque (SOT) with various magnetic materials, we investigated SOT in rare earth-transition metal ferrimagnetic alloys. The harmonic Hall measurements were performed in Pt/GdFeCo bilayers to quantify the effective fields resulting from the SOT. It is found that the damping-like torque rapidly increases near the magnetization compensation temperature $T_\mathrm{M}$ of the GdFeCo, which is attributed to the reduction of the net magnetic moment.

Deluge of the electronic information urges faster, denser, and more energy efficient magnetic data storage applications [1]. Low energy and fast manipulation of the magnetization is the key to realize such devices. Vigorous study on spin transfer torque has already proved that a direct interaction between the magnetization and the electron spins can provide much more energy efficient magnetization control comparing with the traditional Ampere field [2]. It has been recently shown that the spin orbit torque (SOT) originating from the spin-orbit interaction can also manipulate the magnetization in as simple as a heavy metal/magnetic bilayer system [3,4]. While a complete physical understanding of SOT is still on the way [5], search for materials as well as material combinations exhibiting a large SOT is becoming more and more active amid demand for aforementioned applications. Pt, Ta, and W are among the heavy metals exhibiting a large SOT [3,4,6]. On the other hand, materials search for magnetic layers has also recently been extended from ferromagnets to ferrimagnets [7,8] and ultimately antiferromagnets [9,10], in which the antiferromagnetically coupled magnetic moments are expected to be beneficial for low energy and fast magnetization manipulation.

Rare earth (RE)-transition (TM) metal ferrimagnetic alloys are one of the best materials for testing the antiferromagnetically coupled systems with SOT. In the $Gd_{25}Fe_{65.6}Co_{9.4}$ alloy used in the present study, Gd moment and FeCo moment are antiferromagnetically coupled and the difference of the Gd and FeCo moments results in a small net magnetic moment. Like other ferrimagnetic materials, two magnetization dynamic modes coexist [11]; ferromagnetic (slow mode) and antiferromagnetic (fast mode) mode the latter of which may be beneficial for the fast magnetization manipulation. The most interesting property of the GdFeCo is that its net magnetization $M_s$ can be controlled by temperature. Due to the competition between the temperature dependences of Gd and Co magnetic moment and their gyromagnetic ratio, there are two critical temperatures: One is the magnetization compensation temperature $T_M$ at which the Gd and CoFe magnetic moments $M_{s,Gd}$ and $M_{s,CoFe}$ are cancelled each other resulting in no net magnetic moments. The other is the angular momentum compensation temperature $T_A$ at which the angular momentum $M_{s,CoFe}/\gamma_{CoFe}$ and $M_{s,Gd}/\gamma_{Gd}$ are cancelled each other resulting in no net angular momentum. These two compensation temperatures are generally different because the gyromagnetic ratio differs for Gd ($\gamma_{CoFe}\sim 2$) and CoFe ($\gamma_{CoFe}\sim 2.2$) [12,13]. As explained in the following paragraph, these magnetic properties unique to RE-TM ferrimagnets combined with SOT can lead to low energy and, possibly, fast manipulation of the magnetization.

The current density $J_e$ flowing in the heavy metal layer gives rise to two different components of SOT on the magnetization: the "damping-like" torque $\tau_{DL}$ and the "field-like" torque $\tau_{FL}$ respectively expressed as,

$$\boldsymbol{\tau_{DL}} = -\frac{\gamma \xi_{SH} J_e}{M_s t}\frac{\hbar}{2e}(\widehat{\boldsymbol{m}}\times(\widehat{\boldsymbol{\sigma}}\times\widehat{\boldsymbol{m}})) \quad , \quad (1)$$

$$\boldsymbol{\tau_{FL}} = \gamma \beta J_e (\widehat{\boldsymbol{\sigma}}\times\widehat{\boldsymbol{m}}) \quad , \quad (2)$$

where $\hbar$ is reduced Plank constant and $e$ is the elementally charge of an electron. $\widehat{\boldsymbol{m}}$ is the unit vector along the magnetization, $\hat{\sigma}$ is the polarization of electron spins generated from the spin-orbit interaction, and $\gamma$ is the gyromagnetic ratio. $t$ is the thickness of the magnetic layer. $\xi_{SH}$ and $\beta$ are respectively the spin Hall efficiency, containing the interfacial mixing conductance and spin diffusion length, and the coefficient related to the Rashba effect [14]. The damping-like torque, which is generally much larger than the field-like torque and is the main driving force of the magnetization switching, is inversely proportional to $M_s/\gamma$ as is generally explained by the angular momentum conservation. Therefore, a strategy to increase the damping-like torque has been considered reducing $M_s/\gamma$ of the system [15,16]. Increasing the damping-like torque leads to more energy efficient magnetization switching.

In this letter, we investigate the SOT effective fields in a Pt/GdFeCo bilayer in elevated temperature and across $T_M$. By the harmonic Hall measurement [17,18,19], we obtain two kinds of the SOT effective field $H_{DL}$ and $H_{FL}$ respectively resulting from the damping-like and field-like torque. The temperature dependence of these effective fields can translate into $M_s$ dependence of the SOT in the ferrimagnetic system.

The sample structure and the measurement setup with the coordinate system are shown in Fig. 1(a). The multilayers of SiN(5 nm)/Pt(5 nm)/Gd$_{25}$Fe$_{65.6}$Co$_{9.4}$(10 nm)/SiN(5 nm) are formed on a Si substrate by magnetron sputtering. The magnetic easy axis of the GdFeCo is confirmed to be perpendicular to the plane by the magnetization measurement. The film is patterned into a Hall-bar structure with a 5 μm-wide and 90 μm-long channel by a photolithography followed by an Ar ion milling. The dc Hall measurement with an external magnetic field swept in *z* direction show a square hysteresis loop as plotted in Fig. 1 (b), indicating that the patterned Pt/GdFeCo maintains the perpendicular magnetic anisotropy and the Hall voltage is from the anomalous Hall effect (AHE) of the GdFeCo magnetization. While the net magnetization responds to the direction of the magnetic field, the electron scattering responsible for the AHE is known to occur predominantly with the FeCo magnetic moment in GdFeCo system [20]. The reversal of the hysteresis loops at 119 K, as is indicated in Fig. 1 (c), essentially indicates that the $T_M$ occurs at that temperature. [11].

In order to quantify the SOT effective fields in the Pt/GdFeCo layer, the harmonic Hall measurement was employed. We flow an ac current $I_{ac}$ with various amplitudes at 337 *Hz* and measure the Hall voltage in the fundamental ($V_{1\omega}$) and the second ($V_{2\omega}$) harmonic. At the driving frequency of 337Hz sufficiently slower than the ferromagnetic resonant dynamics (~GHz) of the magnetization, SOTs act as an effective field and the magnetization sways in the effective field direction, where the effective fields $\mathbf{H_{DL}} = (\boldsymbol{\tau_{DL}} \times \hat{\boldsymbol{m}})/\gamma = \frac{\theta_{SH}J_e}{M_s t}\frac{\hbar}{2e}(\hat{\boldsymbol{\sigma}} \times \hat{\boldsymbol{m}})$ and $\mathbf{H_{FL}} = (\boldsymbol{\tau_{FL}} \times \hat{\boldsymbol{m}})/\gamma = \beta J_e \hat{\boldsymbol{\sigma}}$. These effective fields are estimated by the response of the magnetization direction to the ac current via the Hall voltage of the fundamental and the second harmonic [18,19]. The magnitude of the effective fields are quantified by,

$$H_{DL(FL)}(H_{x(y)}) = R^{2\omega}\frac{R^{\omega}(H_{x(y)})}{R_{AHE}}(\frac{dR^{\omega}}{dH_{x(y)}}\bigg|_{\theta_0})^{-1} \quad , \tag{3}$$

where $\theta_0 = acos|R^{\omega}(H_{x(y)})/R_{AHE}|$, $R^{\omega} = V_{\omega}/I_{ac}$ and $R^{2\omega} = V_{2\omega}/I_{ac}$ are the fundamental and the second harmonic Hall resistance, $R_{AHE}$ is the amplitude of the anomalous Hall resistance, and $H_{x(y)}$ is the external fields in *x (y)* direction. Equation 3 neglects the planar Hall effect contribution as it is negligibly small comparing to the AHE in our devices (see Fig. (c)). Together with the external field $H_{x(y)}$, a small constant magnetic field (~3 mT) in *z* direction was applied in order to avoid a multi-domain formation. Measurements were performed at the various temperatures from 30 K to 290 K.

Figures 2(a) and (b) show $V_\omega$ and $V_{2\omega}$ as a function of $H_x$ and $H_y$ at the temperature above (250 K) and below (70 K) the $T_M$. The anomalous Nernst-Ettingshausen effect contribution (at the best ~3% of the measured $V_{2\omega}$) was corrected by the technique described elsewhere [19]. Therefore, $V_{2\omega}$ shown here is predominantly from the AHE. The $V_\omega$ curves are due to the AHE from which the magnetization orientation as a function of the external field is deduced. As is shown in Eq. 3, non-zero $V_{2\omega}$ suggests the existence of the SOT effective field. In all the field range, $V_{2\omega}$ with $H_y$ is quite small comparing with that with $H_x$, indicating that $H_{FL}$ is negligibly small comparing to $H_{DL}$. We therefore, only focus on $H_{DL}$ in the following discussion. Figures 2(c) and (d) plot the $H_{DL}$ as a function of $H_x$ calculated by Eq. 3. $H_{DL}$ shows a small variation with $H_x$, suggesting that it depends on the magnetization orientation as reported in Ref. [19]. Here, we only consider $H_{DL}$ at $H_x = 0$ by fitting the data with $H_{DL}(H_x) = H_{DL}(0) + \frac{\partial H_{DL}}{\partial H_x}\big|_{H_x=0} H_x + \frac{1}{2}\frac{\partial^2 H_{DL}}{\partial H_x^2}\big|_{H_x=0} H_x^2 \cdots$ as shown in Figs. 2(b) and

(c).

Figure 3(a) shows $H_{DL}(0)$ as a function of the ac current densities $J_e$ flowing in Pt [21]. $H_{DL}(0)$ increases linearly with $J_e$ but the slope varies with temperature. $H_{DL}(0)/J_e$ obtained by the linear fitting of the Fig. 3(a) is found to show a rapid enhancement as the temperature approaches $T_M$ as shown in Fig. 3 (b).

In the framework of the spin Hall effect, which is considered the main source of the damping-like torque, and of the conventional spin transfer theory having been used for ferromagnets, $H_{DL}(0)/J_e$ is given by the transform of Eq. 1 as,

$$\frac{H_{DL}(0)}{J_e} = \frac{\hbar}{2e} \frac{\xi_{SH}}{M_s t_{GdFeCo}}, \qquad (4)$$

where $t_{GdFeCo}$ is the GdFeCo layer thickness. The equation essentially suggests that the enhancement of $H_{DL}(0)/J_e$ can possibly be due to the reduction of $M_S$ at around $T_M$. As shown in Fig. 3(b), however, $M_S$ obtained by the magnetization measurement of the blanket film did not match the trend Eq. 4 would suggest because of the loss of the Gd moment during the device patterning process. We identified, in separate experiments on the device size dependence, that the Gd moment is indeed reduced at the device edge possibly due to a preferential oxidation of the Gd. Consequently, $T_M$ of the patterned device ($T_M$ = 119 K) is significantly lower than that of the blanket film ($T_M$ = 201 K). In order to compensate the difference in $T_M$, with an assumption that the temperature dependence of $M_s$ at around $T_M$ is the same for the patterned device and the blanket film, we simply shift the temperature vs. $M_s$ curve of the blanket film to $T_M$ = 119 K of the patterned device as shown in the inset of the Fig 3(c). By using the shifted data ($M_{s, shifted}$), we draw the graph of $H_{DL}(0)/J_e$ as a function of 1/ $M_{s, shifted}$ as shown in Fig. 3(c). The $H_{DL}(0)/J_e$ and 1/ $M_{s, shifted}$ are found to have a nice linear relation as expected from Eq. 4. We obtained $\xi_{SH}$= 0.014 $\pm$ 0.002. This result essentially dictates that the spin transfer responsible for the damping-like torque in RE-TM ferrimagnetic GdFeCo simply scales with the reciprocal of the net magnetization. Moreover, it infers that antiferromagnetically coupled Gd and FeCo does not seem to give any peculiarities and the spin transfer in the ferrimagnetic system can be simply explained in terms of the total net magnetization, similarly to that in ferromagnetic systems.

In summary, we investigate SOT effective fields in the Pt/GdFeCo bilayers by the harmonic Hall measurement. $H_{DL}(0)/J_e$ originating from the damping-like torque shows a strong temperature dependence and increases near $T_M$, while the field-like torque is negligibly small in all the temperature range. The $H_{DL}(0)/J_e$ enhancement is found to be due to the reduction of net $M_s$ of the ferrimagnet and the estimated spin Hall efficiency $\xi_{SH}$ is 0.014±0.002. Our results suggest that the spin transfer for the damping-like torque is simply acting on the net $M_s$ of the ferrimagnet. As we only encountered $T_M$ in this study and discussed SOTs around $T_M$, the similar study around $T_A$ should also be interesting.

**Acknowledgement**

This work was partly supported by JSPS KAKENHI Grant Numbers 15H05702, 16H04487,

**Figure captions**

Figure 1 (a) Schematic illustration of the Pt/GdFeCo film structure and the harmonic Hall voltage measurement. $H_{DL}$ and $H_{FL}$ represent the effective fields respectively corresponding to the damping-like and field-like torque. $I_{ac}$ denotes an ac current. (b) The Hall resistances $R_{yx}$ as a function of external magnetic field in $z$ direction $H_z$ at various temperatures. (c) The full change of the Hall resistance $\Delta R_{yx}$ as indicated in (b), absolute value of $\Delta R_{yx}$, and planar Hall resistance (solid red) as a function temperature.

Figure 2 $V_\omega$ and $V_{2\omega}$ at 250 K (a) and 70 K (b). Red curves are $V_\omega$ as functions of both $H_x$ and $H_y$ (they are exactly overlapped). Blue curves are $V_{2\omega}$ as a function of $H_x$ and yellow curves are $V_{2\omega}$ as a function of $H_y$. $V_{2\omega}$ of the yellow curves is multiplied by a factor of five. $H_{DL}$, calculated with Eq. 3 as a function of $H_x$ at 250 K (c) and 70 K (d). The experimental data points are fitted by $H_{DL}(H_x) = H_{DL}(0) + \frac{\partial H_{DL}}{\partial H_x}\Big|_{H_x=0} H_x + \frac{1}{2}\frac{\partial^2 H_{DL}}{\partial H_x^2}\Big|_{H_x=0} H_x^2 \cdots$.

Figure 3 (a) $H_{DL}(0)$ as a function of current densities $J_e$ in the Pt layer at various temperatures. (b) $H_{DL}(0)/J_e$ (markers) and $1/M_{s,film}$ (dashed lines) as a function of temperature. (c) $H_{DL}(0)/J_e$ as a function of $1/M_{s,shifted}$ and the fitted slope (red line). The inset shows $H_{DL}(0)/J_e$ and $1/M_{s,film}$ with normalized temperature ($\Delta T=T-T_M$). Note that index of each axis in the inset is same as Fig 3 (b).

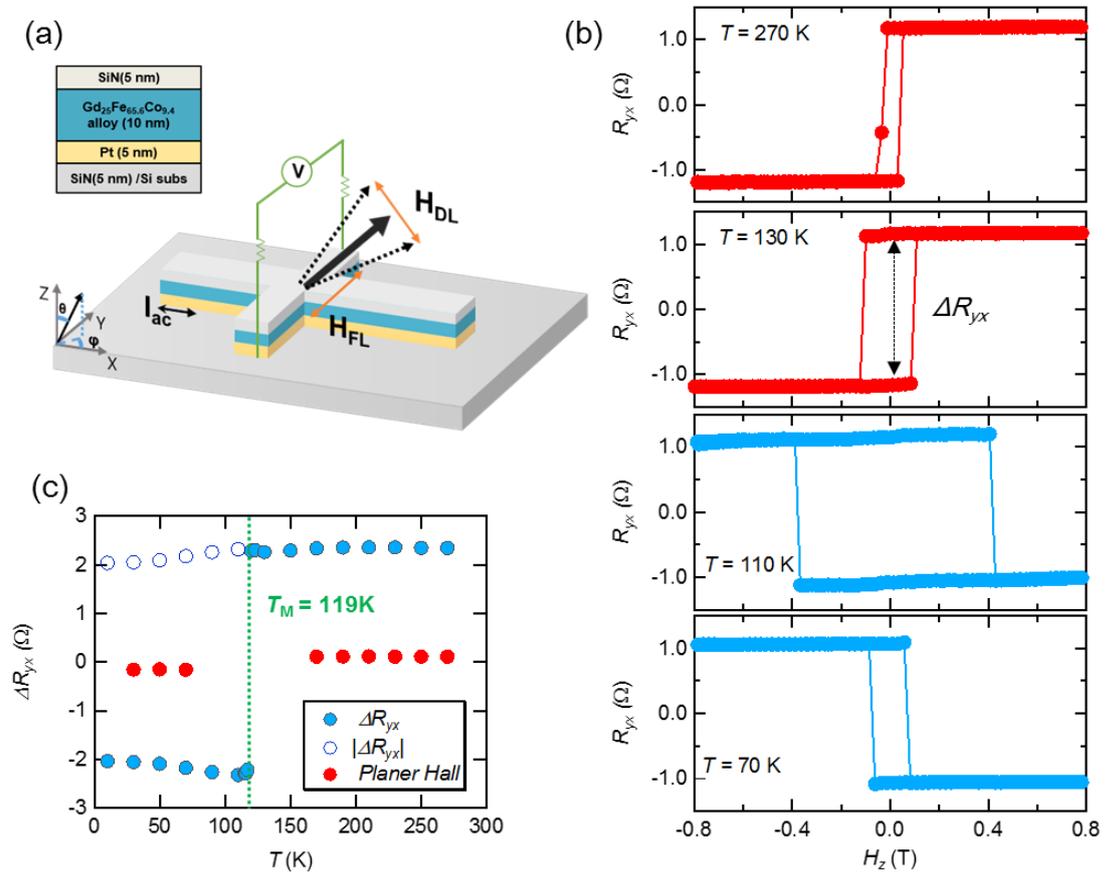

Figure 1 Ham et al.

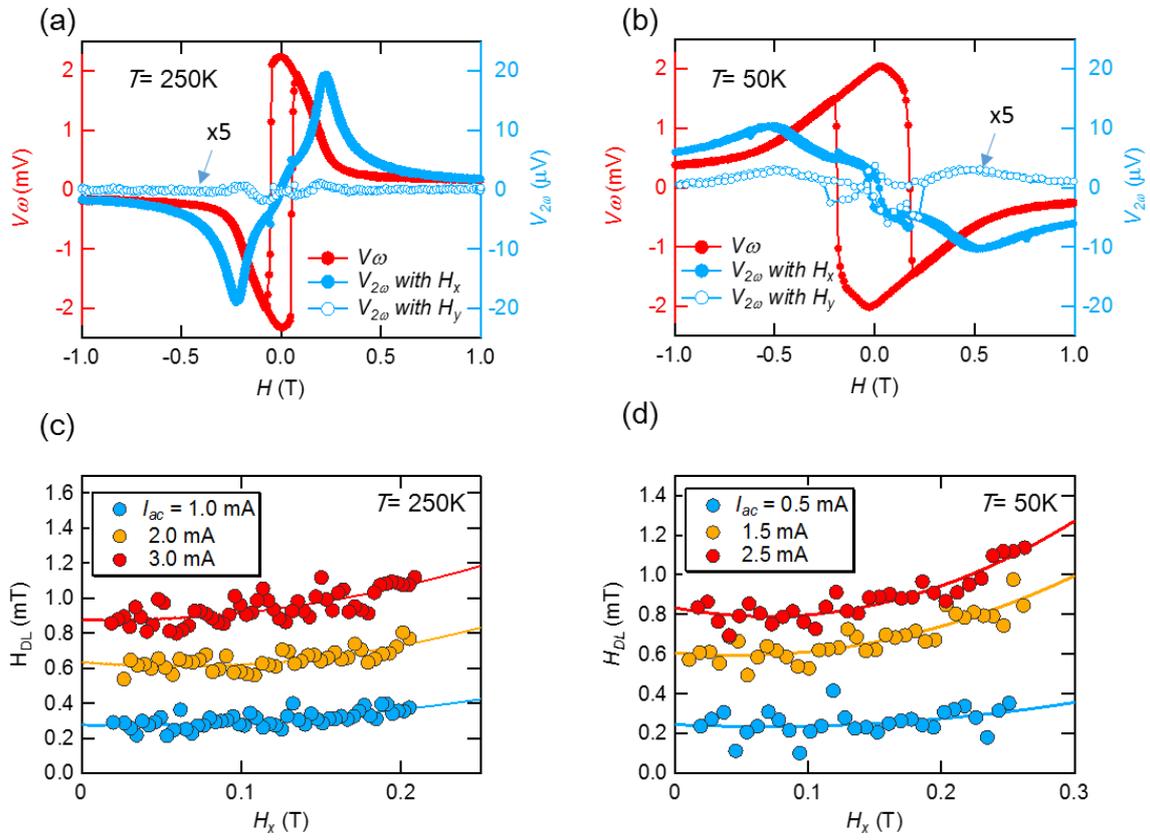

Figure 2 Ham et al.

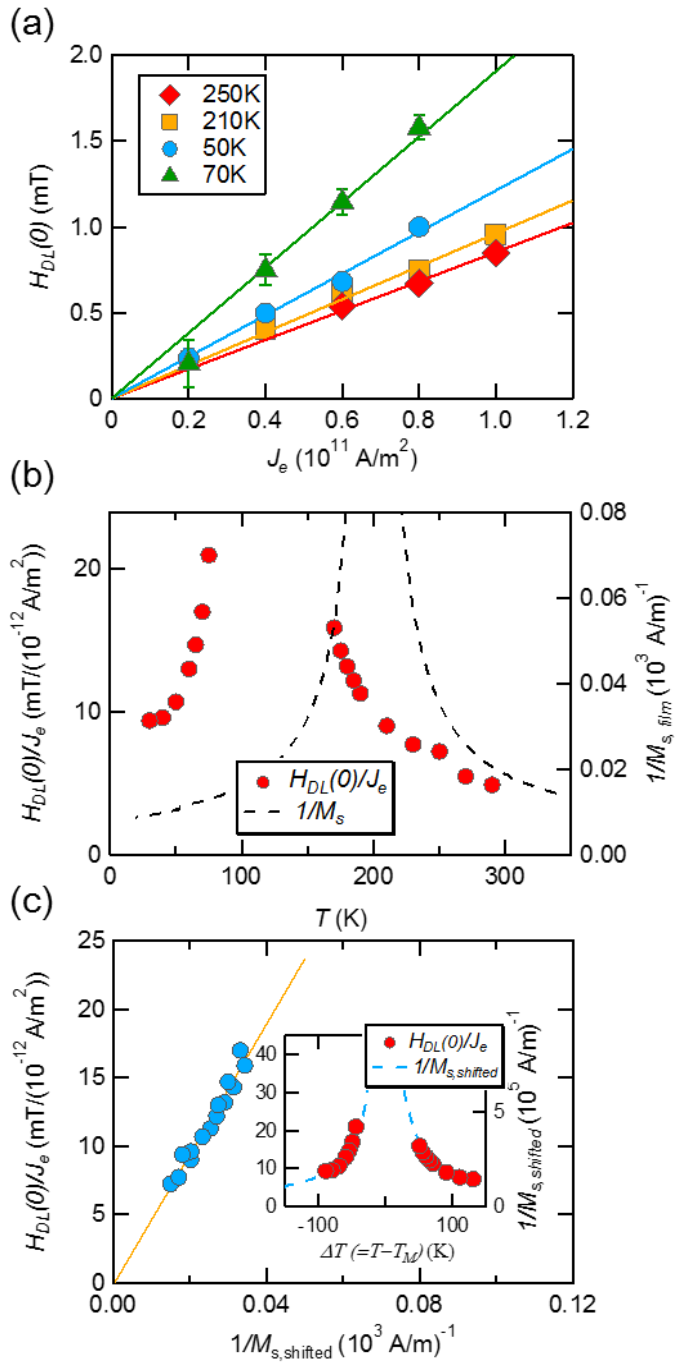

Figure 3 Ham et al.